# Catalyzing Social Interactions in Mixed Reality using ML Recommendation Systems

**Sparsh Srivastava**
ss6381@columbia.edu

**Rohan Arora**
ra3091@columbia.edu

**Department of Computer Science**
*Columbia University in the City of New York*
New York City, NY, 10025

## ABSTRACT

We create an innovative mixed reality-first social recommendation model, utilizing features uniquely collected through mixed reality (MR) systems to promote social interaction, such as gaze recognition, proximity, noise level, congestion level, and conversational intensity. We further extend these models to include right-time features to deliver timely notifications. We measure performance metrics across various models by creating a new intersection of user features, MR features, and right-time features. We create four model types trained on different combinations of the feature classes, where we compare the baseline model trained on the class of user features against the models trained on MR features, right-time features, and a combination of all of the feature classes. Due to limitations in data collection and cost, we observe performance degradation in the right-time, mixed reality, and combination models. Despite these challenges, we introduce optimizations to improve accuracy across all models by over 14 percentage points, where the best performing model achieved 24% greater accuracy.

## 1 INTRODUCTION

Due to recent advancements in mixed reality (MR) hardware, we see that modern MR devices are able to collect a new class of visual data in real-time [9, 10]. These advancements in technology present a pressing need for on-line machine learning models and algorithms that utilize this new swath of data.

### 1.1 Value to user community

The 2023 U.S. Surgeon General's advisory highlighted loneliness as a critical and enduring public health crisis [1, 2, 3]. This work on social recommendations in mixed reality provides value to this user community, composed of lonely individuals located in the United States, ranging from 18 to 65 in age.

**Contributions:**

(1) Focus group analyzing user preferences and present-day perceptions around social interactions with co-located strangers.
(2) Novel dataset on social preference collected through an empirical study conducted using human participants.
(3) Four user-to-user recommendation model types trained on combinations of user features, mixed reality features, and right-time features for predicting missed in-person interactions.

Since this research assumes technological advancements in mixed reality, such as eye-tracking sensors for gaze recognition, front-facing cameras to capture environmental information, etc., the contributions mentioned above may not provide immediate value to the target user group, specifically with regards to mixed reality. The non-MR models provided through this research would deliver immediate value by catalyzing novel social interactions through recommendations on the Hear application platform [2].

### 1.2 Research questions

We plan to address various research questions through the exploration of MR features, right-time features, and user features as inputs to the recommendation models introduced in this paper. The questions answered by this work are captured below:

*RQ1. What is the impact of mixed reality in catalyzing novel social interactions between people within close proximity?*

*RQ2. What environmental factors play the largest role in creating or preventing social interactions?*

*RQ3. What are the primary user features that make people want to interact with each other?*

We describe these feature classes as disjoint sets of attributes based on how we expect that they will be collected. The user features are unique data points of personal information provided by the users. Due to the nature of making unidirectional predictions and recommendations, we expect each set of user features to include values for "self" (i.e. the person receiving the recommendation) and "candidate" (i.e. the person being recommended). The right-time features represent any environmental data around the "self" and "candidate" users that can be collected using traditional devices. The MR features are any visual data that can only be collected by a mixed reality device.

## 1.3 Assumptions

Potential issues with MR-first social interactions between co-located strangers include incomplete or missing information when both users are in a shared physical space, where user A may be visually occluded from user B. We navigate this situation in our scenarios by introducing a friend who notifies user A of user B's presence, without needing to establish visual contact.

Another concern is that only one or neither of the users are wearing MR devices. Due to this potential platform incongruence, we present multiple models which require different subsets of features, where users not wearing MR devices can still receive recommendations and be recommended to other users with however much information is presently available. We make no assumptions on the platform on which the user has set up their profile, but we enforce that the values for MR features in specific were collected using a future MR device capable of providing such information.

## 2 RELATED WORK

We will review the literature across multiple fields, such as human-computer interaction, social psychology, mixed reality, and machine learning. We began our research by conducting an initial focus group with ten participants to inform our future surveys and research methodology. Then, we launched the user study, composed of the user survey and scenario survey, where we qualitatively recorded explanations provided by the participants on which aspects of each scenario impacted their decisions most.

Research in virtual and mixed reality has seen substantial growth across various domains. Prior work has focused on the technological aspects of MR systems, particularly advancements in sensors and hardware [9, 20]. This study is unique, as it utilizes MR features to facilitate social interactions. Existing studies on social recommendation systems for connecting strangers [4, 5] have typically centered on traditional contexts without integrating MR capabilities, or focus on technical aspects around object tracking between multiple user devices [19]. These advancements in hardware sensors for MR devices enable the collection of a new class of real-time visual data [9, 10] that can be used for a variety of use cases [6, 7, 8].

People are often intimidated by initiating interactions with others due to fear of social evaluation [11, 12, 13] where people worry that their "attempts to initiate interactions may be rejected or that others may not enjoy the conversation" [4, 16, 17]. These fears of social evaluation are motivated by an uncertainty about other people's interests and what they are willing to talk about [14, 15]. People naturally avoid situations that can cause embarrassment or social awkwardness [12]. We observe that knowledge of shared interests between people enables conversation [17, 18], however, these attributes are not always shared publicly by people. When common interests are not immediately observable, people can feel deterred from initiating interactions with new people due to the fear of social evaluation [12]. We propose a framework for incorporating user interests into their profile, such that our models can make accurate and relevant recommendations.

We make a novel contribution to the field by integrating MR-specific features, such as gaze tracking, occlusion, and congestion level detection, into our dataset. Additionally, the inclusion of right-time features, such as location, weather, and time of day, further enhances the models' understanding of the user's environment. This unique combination of user, MR, and right-time features (as seen in section 3.2) allows for a more holistic approach to understanding and predicting social interactions in mixed reality contexts. Modern recommendation techniques such as collaborative filtering and content-based systems have become widely utilized for providing personalized item-to-user and user-to-user recommendations [4, 21, 22]. Our proposed feature sets allow for a deeper understanding of social dynamics in physical environments, enabling recommendation systems to provide real-time suggestions.

## 3  METHODOLOGY

We conducted a focus group and user study which collected data from real-life participants through two separate surveys. First, we circulated the user survey to participants through the Prolific platform, which we used to collect our user features and ask questions for our focus group. Then, we distributed the scenario survey, where we defined questions with scenarios generated using the user profiles provided in the initial user survey. We used these scenarios to understand the behavior of the survey participants in various social situations.

### 3.1  Focus group

This focus group aims to understand how users establish social proof and their preferences for socialization in an MR context. In addition to collecting user features in the user survey (outlined in section 3.2), we also ask the following questions for our focus group:

- How frequently do you want to interact with someone new when you are in public?
- How easy or difficult is it to meet new people in public?
- How would you prefer to interact with new people for your first interaction?
- Which of the following points of personal information would you selectively share about yourself with strangers in your immediate vicinity? (multiple select)
- How much time do you spend listening to music in public per day?
- Would you want to connect with a stranger who has a similar taste in music as you?
- How useful would it be to receive a notification when someone nearby wants to connect with you?

The collected user information from the focus group survey will not only provide context for user socialization preferences, but will also serve as user data for the generated scenarios of the scenario survey (described in section 3.2). In addition, the participants from the focus group will be the same individuals that participate in the user study.

### 3.2  User study

The user study, composed of the user survey and the scenario survey, captures all of the features required for model training. After acquiring all of the necessary consent, the user survey asks the survey participants to input their profile information through questions like "What is your age?", "what is your hair type?", and "are you currently a student?". We use this information to generate scenarios for the scenario survey where the participants are posed with a situation where they are receiving a recommendation to interact with another user. The scenario survey is sent to the same participants in the user survey, such that each participant answers a scenario question about every other participant in the study, including themselves.

The data collected from this user study enables the prediction of three output classes – "Meet (in person)", "Chat (via instant messaging)", and "Reject". These classes represent the different choices that users can make when they receive a recommendation. The output label has been simplified to these three categories, reflecting insights from our focus group findings, which indicate significant user preference for in-person interactions and interactions via instant messaging.

Consider the sample scenario[1], as described in the scenario survey of the user study, where the study participant is asked to decide whether or not they want to interact with the described user. The underlying assumption is that this user profile is being shown to the survey participant because they have both chosen to publicly share their profile with other users within some local proximity. Scenarios measuring social interactions in mixed reality assume the consent of both parties involved. These scenarios will be presented to users in our study to measure various data points about the underlying intricacies of the future of social interaction.

The collected data will include user features (age, gender, education level, student status, workforce status, industry, favorite hobby, favorite interest, favorite music genre, personality, favorite social media, daily music listen time), MR features (height, hair type, hair color, tattoos, occlusion, proximity, human congestion level, gaze "self" to "candidate", gaze "candidate" to "self", conversational intensity, "self" clothing, "candidate" clothing), and right-time features (location, weather, human noise level, non human noise level, day of the week, time of day). For instance, in the previously mentioned sample scenario, glance serves as a proxy for gaze, which would be recorded by an MR device using eye tracking sensors.

### 3.3 Data preparation

We created a *scenario-builder* script to generate the scenario survey, which required parsing the user survey responses into data structures, selecting non-deterministic values for the MR and right-time features from sets of possible values, and outputting scenarios for specific participants in natural language. Additionally, we validate the responses to ensure that there are no glaring inconsistencies. Given the unique structure of the surveys, participants who fulfill this criteria[2] would be automatically rejected or flagged from the study. The non-deterministic feature values, such as gaze "self" to "candidate", gaze "candidate" to "self", "candidate" occluded, human congestion level, "candidate" conversational intensity, proximity, weather, human noise level, and non human noise level are randomly sampled from the possible values for each of the features every time a new scenario is being generated. Location-based non-deterministic features, such as location, day of the week, time of day, group size, and clothing, are derived using the participant responses in the user survey, where they specify what time they generally attend various locations, how many people they would go with, and what they would wear. By combining the user data for "self" and "candidate" with the non-deterministic features and decision responses recorded in the scenario survey, we define 198 unique data points with 73 features and one target variable, collected from over 40 participants across six different surveys (two user surveys and four scenario surveys) into a single dataframe [23] to be used for model training and validation.

---

[1] See Appendix A.

[2] See Appendix B.

### 3.4 Data distribution

Due to limitations in the data collection process caused by survey costs and scalability issues, our dataset exhibited significant data sparsity across multiple feature categories. We also found that many features in the dataset had constant values, which may have led to challenges in generalization for our predictive models when they encountered data points with feature values previously unseen during training.

In terms of user features, we observe relatively high variance in age for both "self" and "candidate," with a higher concentration around ages 25 to 35.[3] The distribution of gender is slightly skewed towards males for "candidate," while there are almost twice as many males as females for "self."[4] Education is heavily skewed towards individuals whose highest education is attainment of a bachelor's degree for both "self" and "candidate."[5] Finally, the majority of data samples are for introverted "self" and "candidate" individuals, with very few identifying as primarily extroverted.[6]

When considering mixed reality features, the number of scenarios where "candidate" was occluded is approximately equal to those without occlusion. The number of scenarios where "self" was looking at "candidate" (gaze) was also approximately equal to those where "self" was not looking at "candidate." However, there were approximately twice as many scenarios where "candidate" was looking at "self," compared to those where "candidate" was not looking at "self."[7] Although not perfectly distributed, human congestion level and proximity of "self" and "candidate" were represented in scenarios relatively evenly. Finally, the majority of the group sizes for "candidate" were with less than 4 other people.[8]

For right-time features, the majority of scenarios took place on weekdays in the afternoon or evening, and on cloudy or rainy days.[9] Additionally, the majority of scenarios had loud human noise level and music playing (loud non-human noise level).[10]

One critical issue with the data distribution was the imbalance in the output label class. Initially, the dataset aimed to predict three outcome classes: "Reject" (88 samples), "Meet (in-person)" (57 samples), and "Chat (via instant messaging)" (53 samples). To address this imbalance, we combined the two positive outcome classes, "Meet (in-person)" and "Chat (via instant messaging)" into a single "Accept" class with 110 combined samples[11], resulting in a more balanced distribution.

## 4 EVALUATION

After preparing the dataset, we utilized scikit-learn's *RandomForestClassifier* with this parameter grid.[12] Specifically, we implemented a grid search with 5-fold cross validation for training various random forest classifiers. We use random forest instead of a collaborative filtering approach for a more straightforward training process.

For each combination of hyperparameters, two sets of four model types were trained, for a total of eight models. Each set of models included the following:
1. ***Baseline model.*** Includes user features from "self" and "candidate."
2. ***Mixed reality model.*** Includes user features from "self" and "candidate" and mixed reality features.

---

[3] See Figure 1 in Appendix E.
[4] See Figure 2 in Appendix E.
[5] See Figure 3 in Appendix E.
[6] See Figure 4 in Appendix E.
[7] See Figure 5 in Appendix E.
[8] See Figure 6 in Appendix E.
[9] See Figure 7 and Figure 9 in Appendix E.
[10] See Figure 8 in Appendix E.
[11] See Figure 10 in Appendix E.
[12] See Appendix H.

3. ***Right-time model.*** Includes user features from "self" and "candidate" and right-time features.
4. ***Combination model.*** Includes user features from "self" and "candidate," mixed reality features, and right-time features.

The first set of these four model types utilized "Meet (in person)," "Chat (via instant messaging)," and "Reject" as options for the target variable. After preliminary analyses, we attempted to boost performance metrics via means other than hyperparameter tuning. Based on the rationale provided in section 3.3, we combined the two positive classes of "Meet (in person)" and "Chat (via instant messaging)," into a single "Accept" class, which consistently achieved higher performance throughout our tests. Thus, we added a second set of models with updated target variable classes as part of our full model optimization routine.

For each hyperparameter combination in our grid search, we output a single text file containing performance metrics for the eight models built in the iteration. For each model, we output the accuracy, precision, recall, and F1 score. Weighted, micro, and macro metrics were output for each of precision, recall, and F1 score. These metrics were output at each fold of 5-fold cross validation, as well as the average metrics across all folds. In addition, we saved each model built and kept track of the best models according to performance metrics. In total, 36,872 random forest classifiers were built, with the model optimization routine taking approximately 10-12 hours in total.

## 5 RESULTS

We outline rationale explaining the performance metrics for each of the recommendation models, and the focus group results from the survey participants that quantitatively support those explanations.

### 5.1 Model results

After completion of the model optimization routine, a "best" model was selected for each of the eight models described in section 4, based on those that achieved the highest accuracy. The performance of each model is shown in *Figure 1* and *Figure 2*.

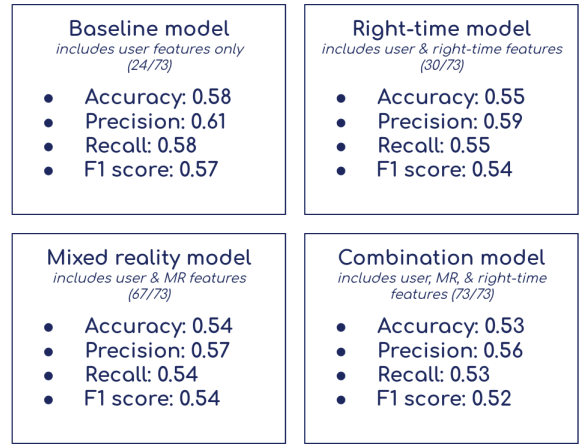

*Figure 1: Accuracy and weighted precision, recall, and F1 scores for three prediction classes - {Meet, Chat, Reject}.*

For the models with three prediction classes (*Figure 1*), the baseline model achieved the highest performance across all metrics, followed by the right-time model, mixed reality model, and finally the combination model (performing the worst).

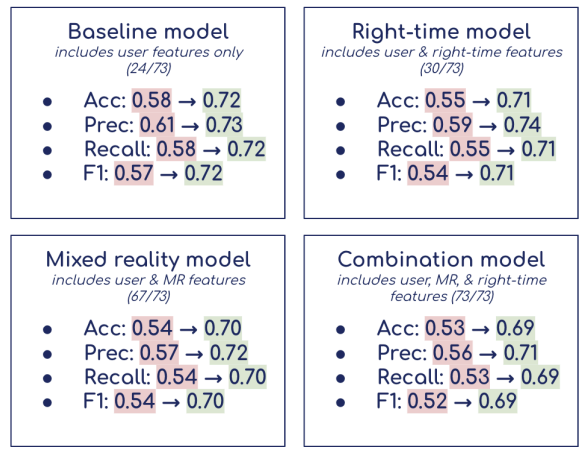

*Figure 2: Accuracy and weighted precision, recall, and F1 scores for two prediction classes - {Accept, Reject} (shown in green).*

With the number of prediction classes reduced from three to two: ({*Meet, Chat, Reject*} → {*Accept,*

*Reject*}), the performance of each model type significantly improved (*Figure 2*). The best performing model type achieved 24% greater accuracy (0.14 increase) with this change. Weighted precision, recall, and F1 score also achieved similar increases. The relative ordering of model types by performance remained consistent compared to the models with three prediction classes, with the exception being that precision was marginally higher for the right-time model compared to the baseline model.

**5.2 Focus group results**

We conducted a focus group of 50 participants over three separate surveys, where we found that approximately 66% of the participants were introverts[13], where the vast majority of the total participants classified the ease or difficulty of meeting new people in public as "hard"[14]. Most people stated that they would want to interact with someone new in public "Somewhat often," but the second most popular choice was "Never." As a means of catalyzing these connections, over 50% of the survey participants stated that they would want to connect with a stranger who has a similar taste in music as them[15].

Most notably, 30 out of the 50 participants stated that they prefer for their first interaction with someone new to be "In-person", and 17 out of 50 participants stated "Instant messaging," where the majority of people would find receiving a notification when someone nearby wants to connect with them to be "Sometimes useful"[16]. In terms of privacy concerns caused by sharing personal information publicly with strangers in your surroundings, the majority of participants stated that they would share their first name / nickname, age, and hobbies, with gender, profile picture, and music taste as other popular choices. However, most people did not choose to share their tattoos, imprecise location, hairstyle, social media, or clothing[17].

Additionally, we asked survey participants to qualitatively explain their decisions and we found that those that wanted to meet were influenced by similar interests, similar age, location, compatibility, and appearance. Those that wanted to chat were influenced by introversion, group size, shared interests, and compatibility. Finally, those that rejected the interaction were influenced by large age gaps, location, group size, differences in interests, and gender.[18]

**6 DISCUSSION**

To address **RQ1**, based on the model performance observed in section 5.1, each of the measured model metrics (accuracy, precision, recall, F1 score) are degraded after the addition of the MR and right-time feature classes as compared to the baseline model type. This degradation may have been caused by the small number of participants involved in the study, which was further compromised by the reduced number of respondents in the scenario survey. The scenario survey required that all of the respondents be participants who had been involved in some version of the user survey, since the profile information presented in the scenario survey was based on real user data. We chose this approach over creating fake profiles because we wanted to record both sides of a social interaction for each participant: one where they receive a notification, and another where they are the user profile that is being recommended. Since only 26 out of the 40 surveyed participants responded to the scenario survey, our data was drastically reduced from the anticipated 400 data points (four rounds with groups of ten participants each) to 198 data points [23]. Additionally, the total number of surveys and participants was restricted by the cost and time to the researchers, due to a limited budget. We observed sparseness for several features in our dataframe, which affected the models' ability to generalize to data not seen during the training process. Since we used 5-fold cross validation, we separated out 20% of the data as the validation set, further reducing the size of our training set. Due to imbalance in the scenario generation process, the data distribution for several classes was heavily skewed, specifically for our non-deterministic features. Many non-deterministic features had missing feature values, which caused the generalization issues mentioned previously. However, these issues would not have substantial impact had it been possible to collect more data.

---

[13] See Figure 1 in Appendix F.
[14] See Figure 2 in Appendix F.
[15] See Figure 4 in Appendix F.
[16] See Figure 3 in Appendix F.
[17] See Figure 5 in Appendix F.
[18] See Appendix G.

We were able to make significant model improvements despite these drawbacks. By combining the two positive classes into a single accept class, we boosted accuracy up to 71% for our highest performing model, with significant improvements to precision, recall, and F1 score as well. This increase in performance could be attributed to the model having fewer output classes to predict between, making it more likely to randomly guess the label correctly. Conversely, the target label values being more evenly distributed could have also influenced this result. Additionally, predicting two output labels is as useful, if not more useful for notifying users, especially given the boost in model performance across the board. This is due to our assumption that users whose decision matches one of the positive target label values are open to connecting with the recommended user regardless of their specific choice.

For **RQ2**, the environmental factors, as encapsulated by the right-time feature class, found self_education level, self_student status, and candidate_workforce status to be the most important features for the {Meet, Chat, Reject} model[19] and self_education level, candidate_education level, and favorite music_genre for the {Accept, Reject}[20] model. For **RQ3**, we observed that the baseline model performed the best out of all the model types. The results of the baseline model[21] were most influenced by candidate_workforce status, self_education level, and favorite music_genre in the {Meet, Chat, Reject} model, and self_education level, favorite music_genre, and candidate_eduction level in the {Accept, Reject} model[22]. According to our data, it appears that a person's level of education, working status, and favorite music genre play an important role in providing social proof of that person's character.

From our focus group results, we identified that the majority of individuals want to interact with someone new when in public, at least somewhat often. In addition, the majority of individuals believe that such interactions are hard or very difficult, with the majority wanting to meet in-person. Given these results, we are encouraged to continue building upon our research to reduce the barriers to social connection in public spaces. Since the majority reported that receiving a notification from nearby individuals wanting to connect would be useful at least some of the time, we would like to extend our work by implementing a real-time notification system which matches users in close proximity. As a follow-up, we would like to verify the focus group results by increasing the number of extroverted participants, as these individuals may offer additional perspectives. We would also like to repeat our study with a larger group of participants to address various data sparsity issues.

---

[19] See Figure 2 in Appendix C.
[20] See Figure 2 in Appendix D.
[21] See Figure 1 in Appendix C.
[22] See Figure 1 in Appendix D.

**ACKNOWLEDGEMENTS**
We thank Professor Gail Kaiser, PhD. for supporting the development of this project. We sincerely thank the participants of this study and the Prolific platform for enabling data collection to train the various models. This work was supported in part by the School of Engineering and Applied Sciences at Columbia University in the City of New York. Any opinions, findings, and conclusions or recommendations expressed in this material are those of the authors and do not necessarily reflect the views of the supporting entities.

**DELIVERY**
Our GitHub repository contains the complete datasets from the user surveys, python scripts for data preparation, and the trained machine learning recommendation models used to trigger social interactions in mixed reality.

https://github.com/ss6381/social-interactions-ml

**APPENDIX A:** *Sample scenario*

You are at your **workplace** on a **weekday** in the **morning**. It is a **cloudy** day and it's **crowded**. There are **many people talking** and **no music is playing**. A friend mentions that they see a person **looking at you** who is **far away**, but you are **unable to see** them from your angle. The person is with **more than 6 other people** and they are **listening to others** in their group.

*You learn the following information about the person:*

**Visual Appearance**
* Height: 6 ft 2 in
* Hair: Black & Short (straight).
* Clothing: Formal clothes.
* Tattoos: None.

**Demographics**
* Gender: Male.
* Age: 22.
* Education: Bachelor's Degree.
* Personality: Introvert.
* Student status: Not currently a student.
* Industry: Transportation & utilities.
* Workforce status: Currently employed.

**Interests & Hobbies**
* Favorite hobby: Games (e.g., board games, video games, puzzles).
* Favorite interest: Personal finance.
* Favorite music genre: Pop (never listens in public).
* Favorite social media: Reddit.

*You and this individual both receive a notification that you two should connect.*

Given the scenario above, would you choose to:
    A. Meet (in person)?
    B. Chat (via instant messaging)?
    C. Reject.

**APPENDIX B:** *Rejection / flagged criteria*

Rejection reasons:
- [REJECTED] they said they go to university frequently, but select later that they don't.
- [REJECTED] they select they don't have a university but they choose clothing for university.
- [REJECTED] their answers in the location-specific questions should be consistent. They should answer "not in university" for all questions or none.
- [REJECTED] they said they go to the workplace frequently, but select later that they don't.
- [REJECTED] they said they are not in the workforce, but select the workplace as a frequently visited location.
- [REJECTED] they select they don't have a workplace but they choose clothing for the workplace.
- [REJECTED] their answers in the location-specific questions should be consistent. they should answer "not in the workplace" for all questions or none.
- [REJECTED] they stated that their furthest in-progress or completed degree was high school or GED, even though we filtered users out by education level.
- [REJECT] they said that their favorite interest is not one of the topics that interest them.

Flag reasons:
- [FLAGGED] they said they are a student, but later that they don't go to university.
- [FLAGGED] they said they are in the workforce, but select later that they don't have a workplace.
- [FLAGGED] did not list university as a frequent location, but answered the location-specific questions for university.
- [FLAGGED] did not list workplace as frequent location, but answered the location-specific questions for workplace.
- [FLAGGED] they said that they were working on the weekends, evenings, or nights.
- [FLAGGED] they said that their favorite hobby is not one of the hobbies that they currently do.
- [FLAGGED] introverts tend to speak more or extroverts tend to listen more.
- [FLAGGED] their favorite social media is not one of the social medias that they use more often.

**APPENDIX C:** *Feature importance for baseline, mixed reality, right-time, and combination models predicting {"Meet", "Chat", "Reject"}.*

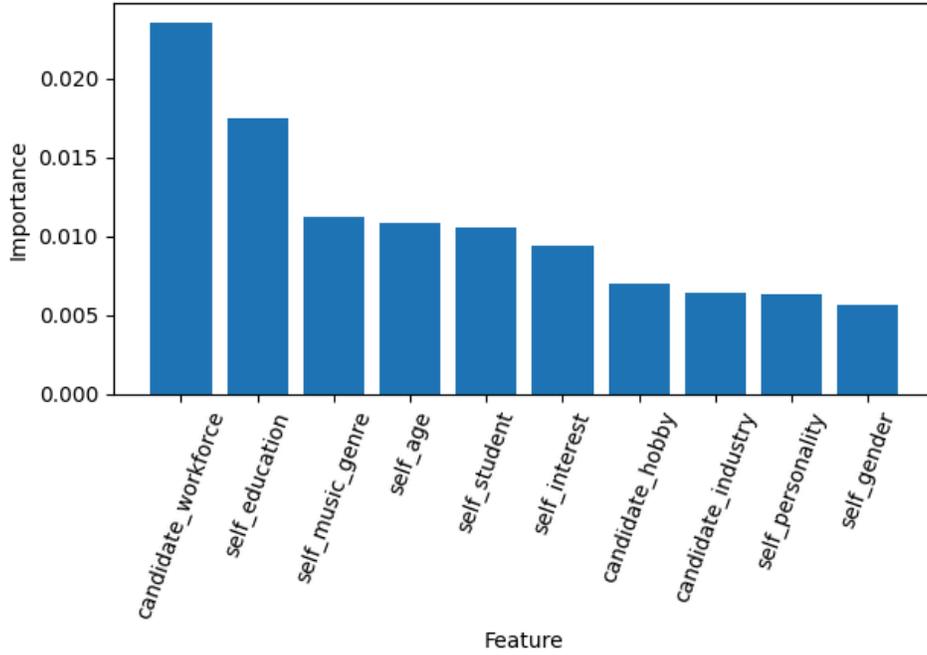

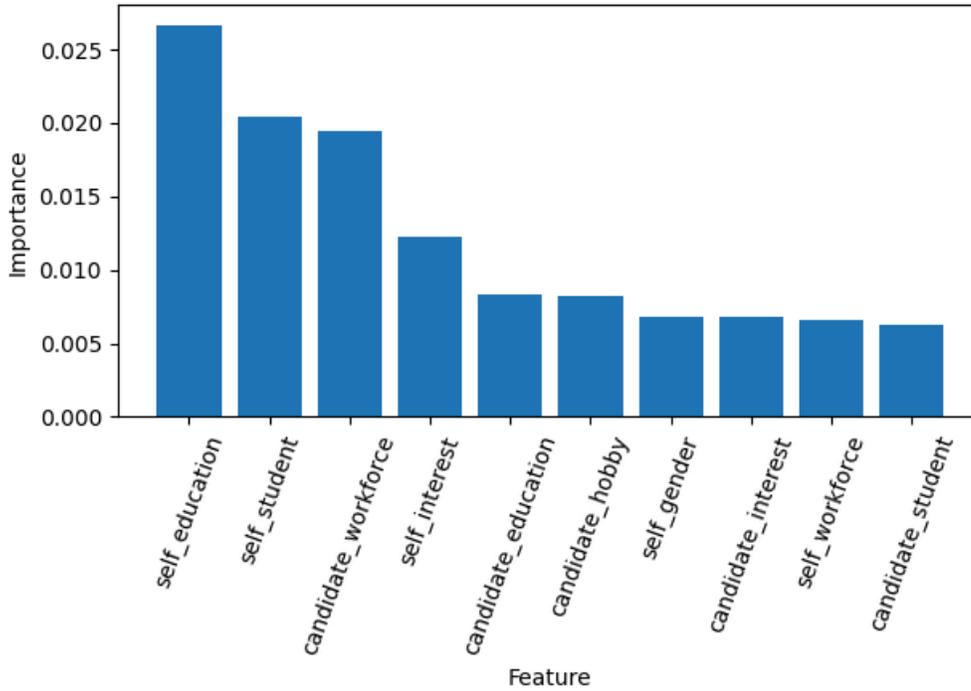

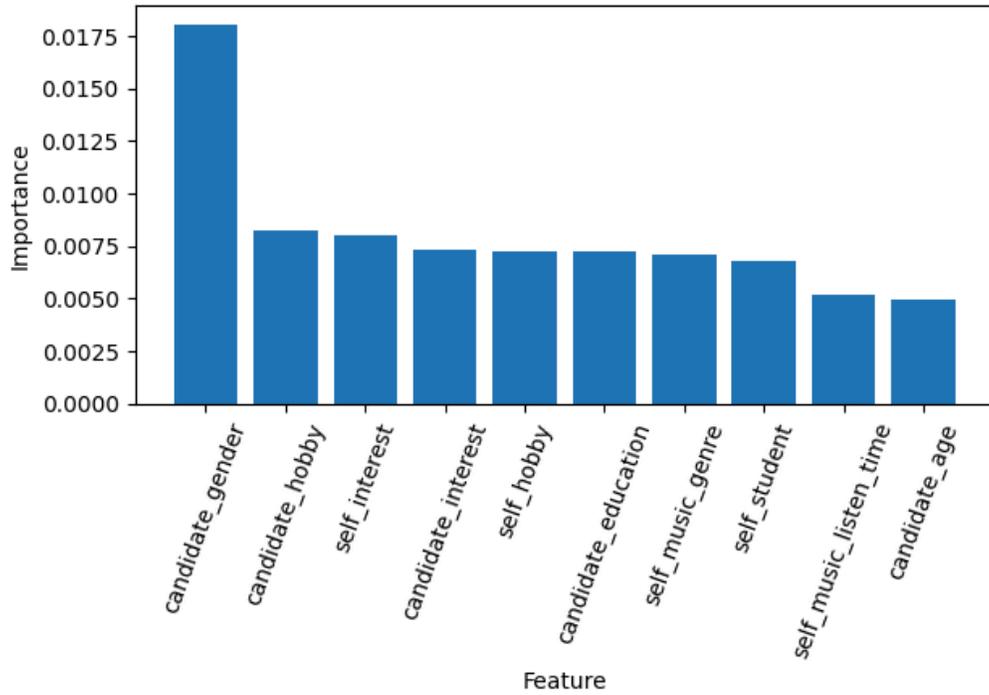

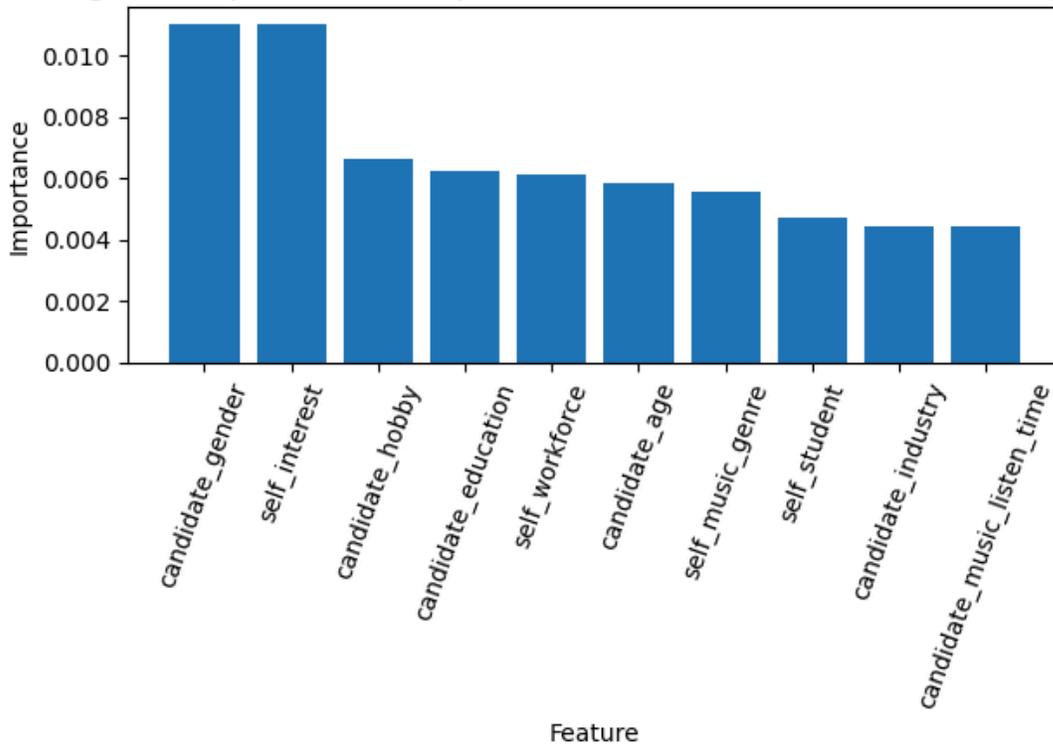

**APPENDIX D:** *Feature importance for baseline, mixed reality, right-time, and combination models predicting {"Accept", "Reject"}.*

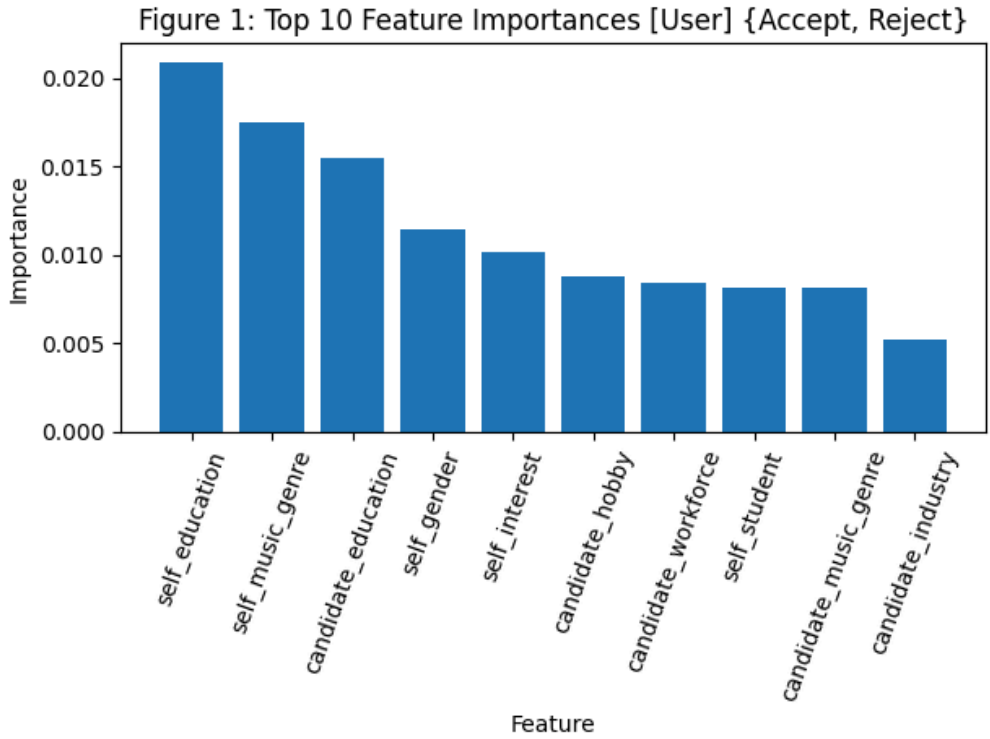

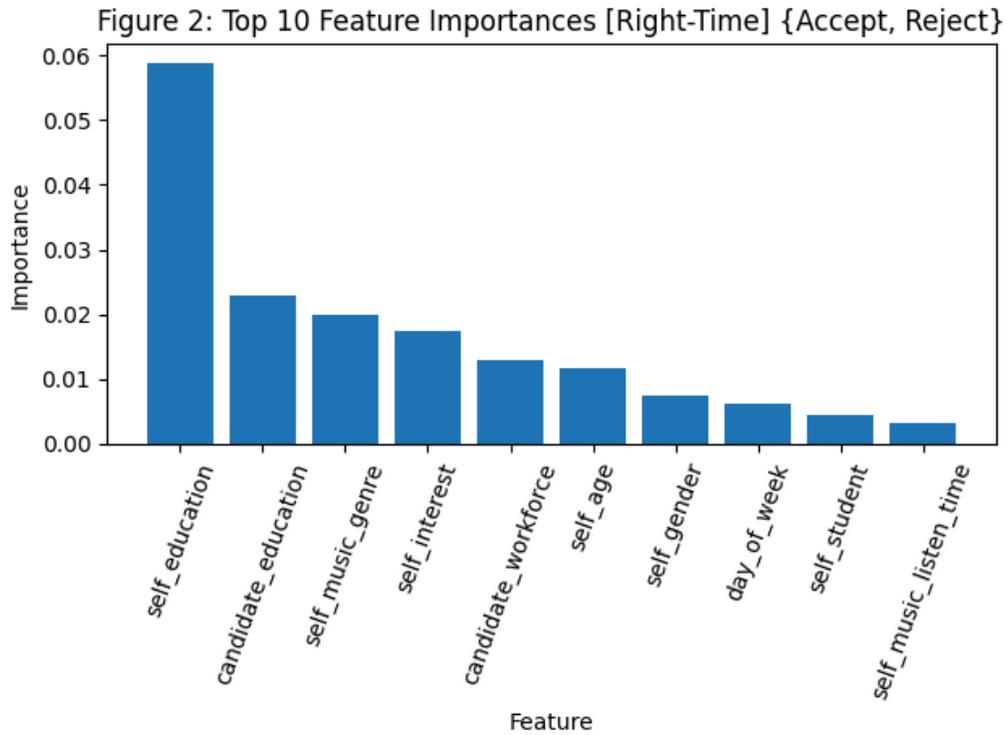

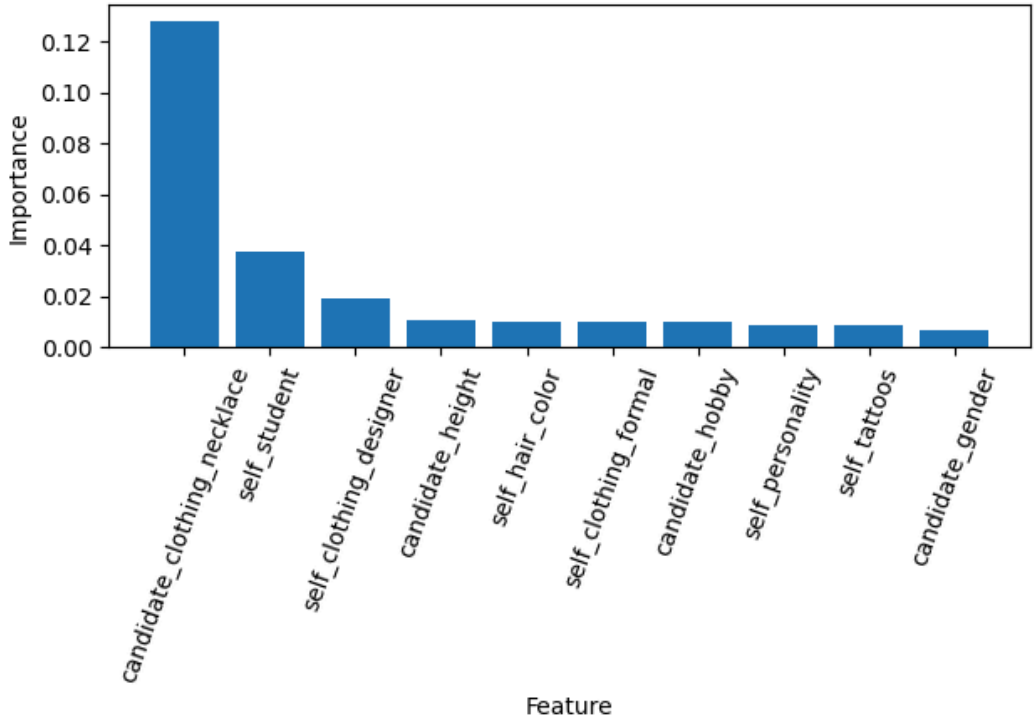

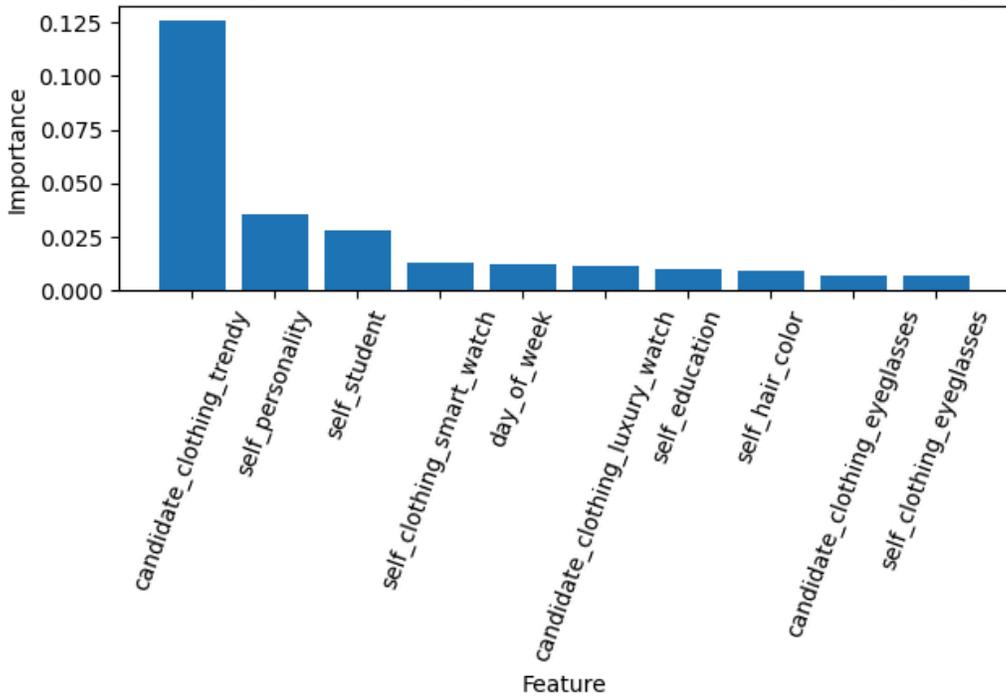

## APPENDIX E: *Distribution of features in the dataset used to train RandomForestClassifier models.*

*Figure 1: [User features] "Self" and "candidate" age.*

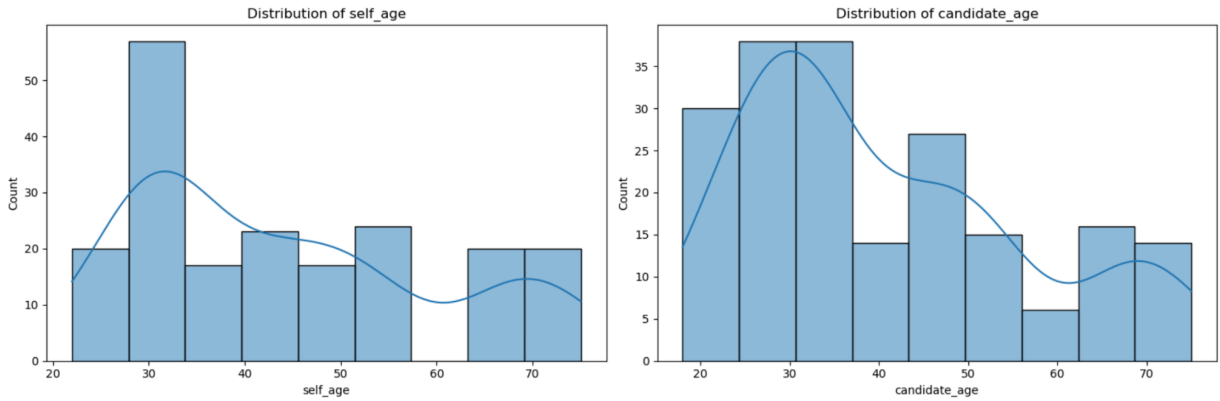

*Figure 2: [User features] "Self" and "candidate" gender.*

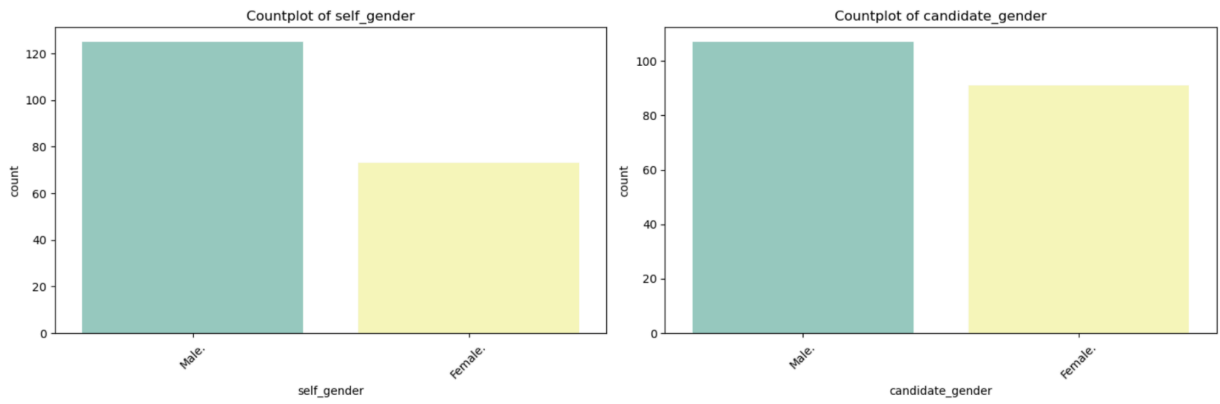

*Figure 3: [User features] "Self" and "candidate" education.*

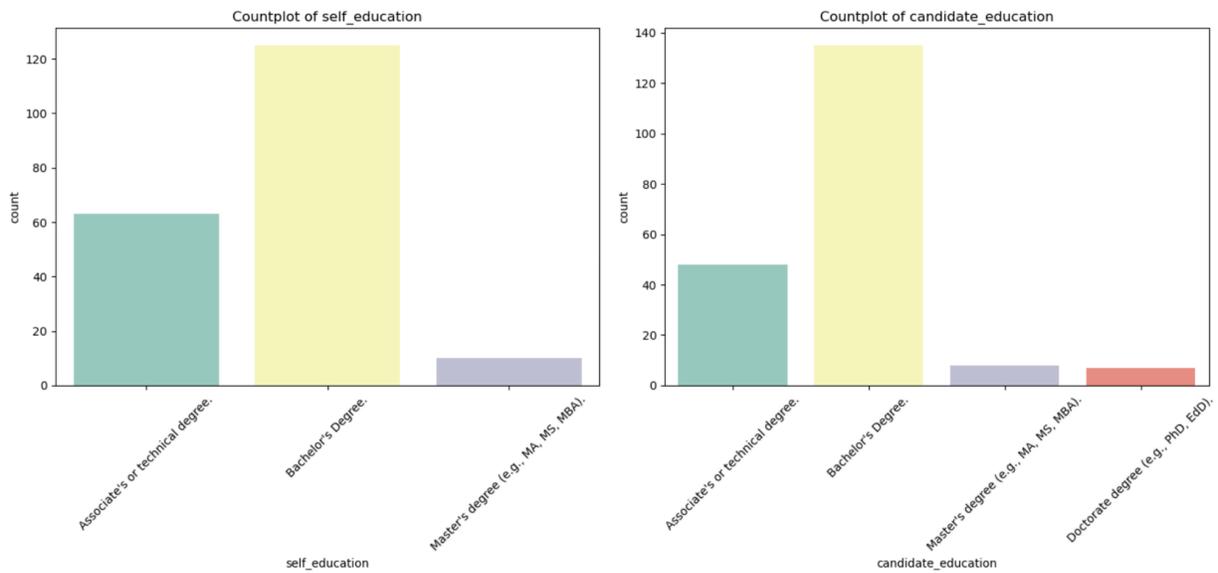

*Figure 4: [User features] "Self" and "candidate" personality.*

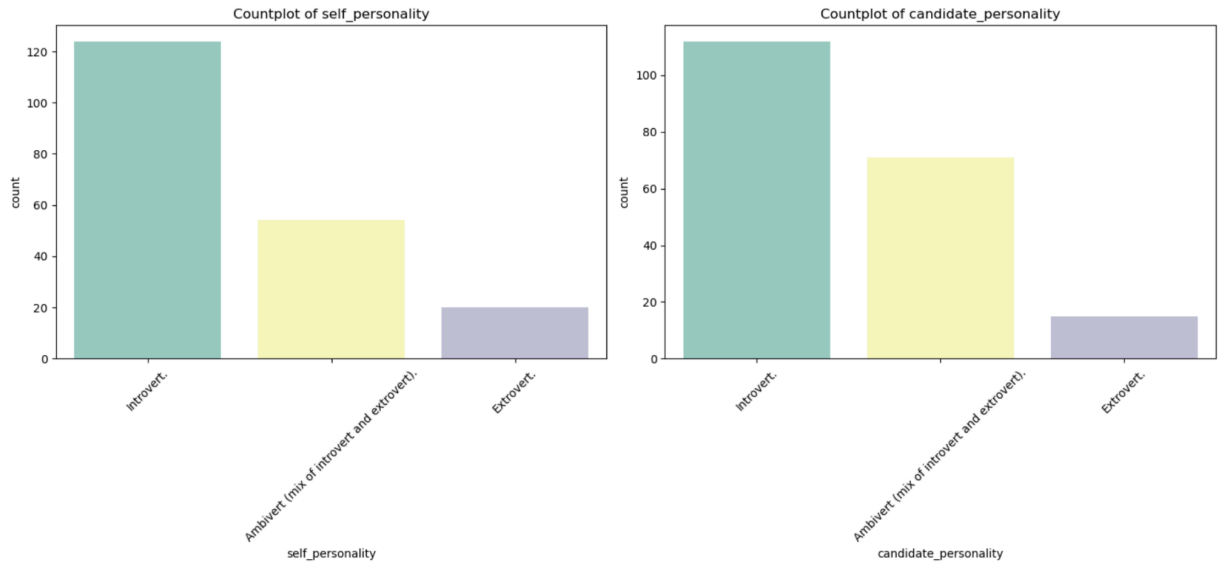

*Figure 5: [Mixed reality features] Occlusion and gaze of "self" and "candidate."*

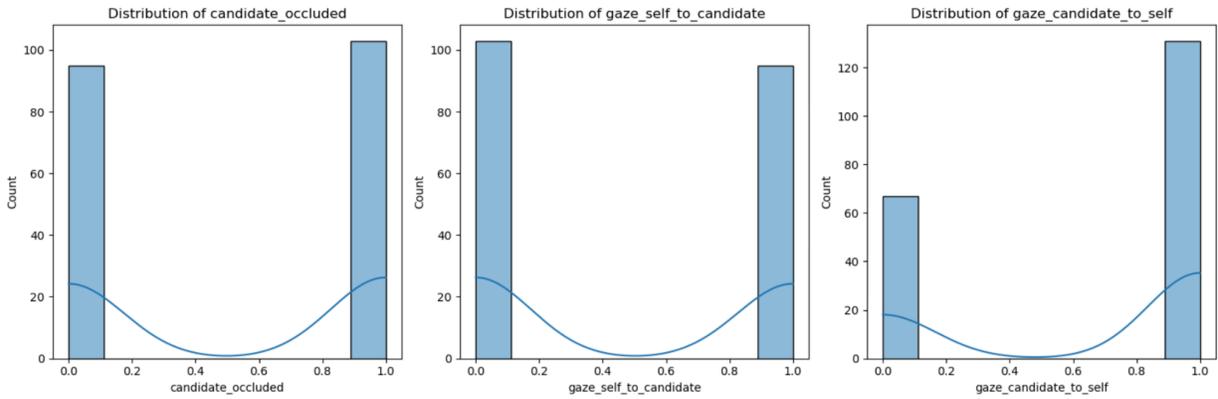

*Figure 6: [Mixed reality features] Human congestion level, proximity, and "candidate" group size.*

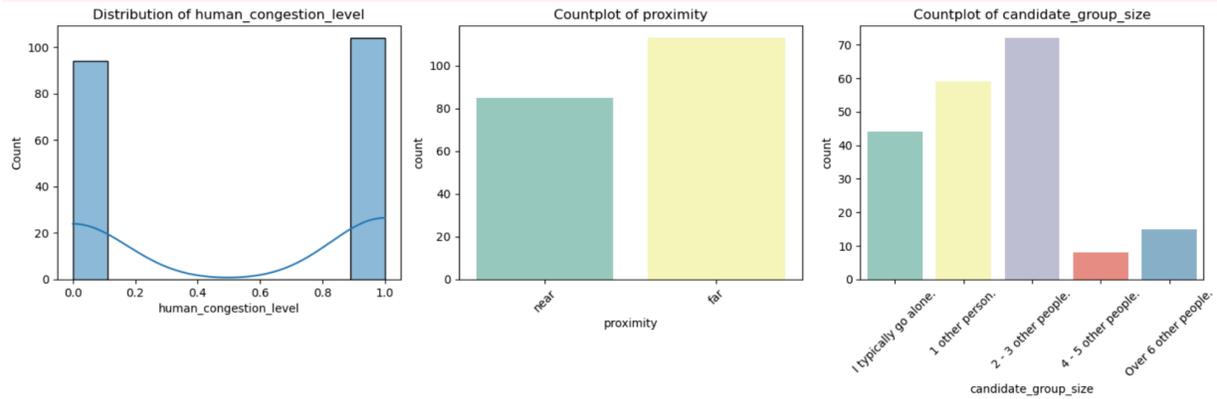

*Figure 7: [Right-time features] Day of week and time of day.*

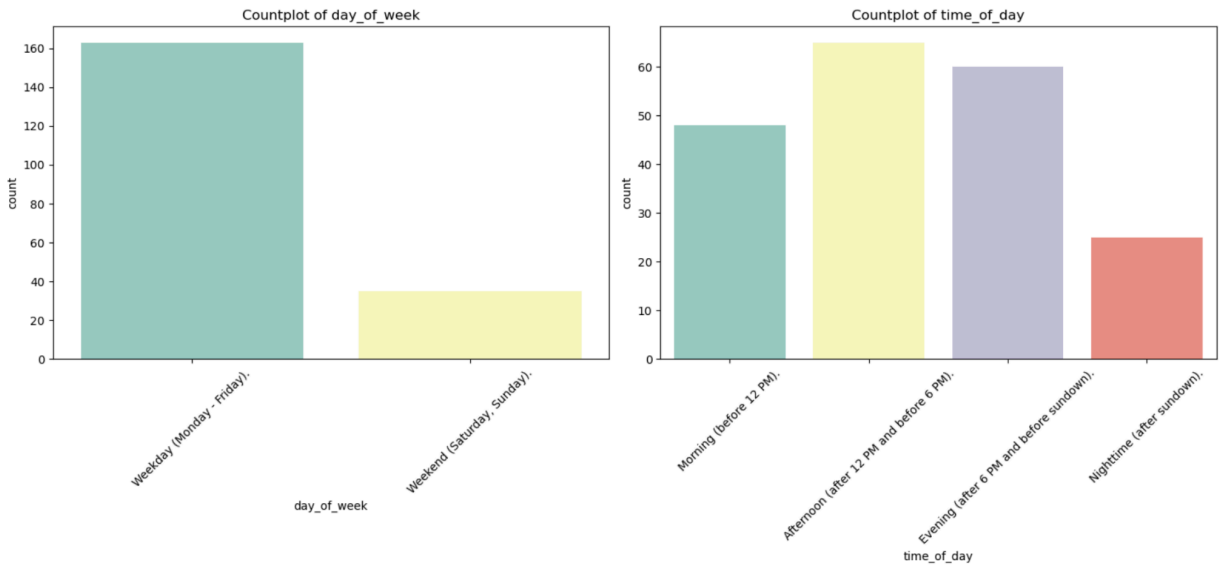

*Figure 8: [Right-time features] Human noise level and non-human noise level.*

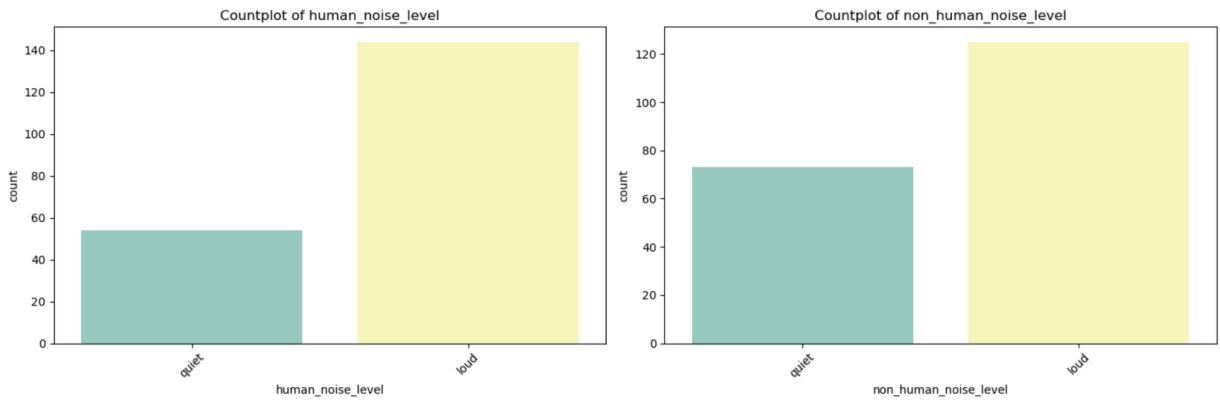

*Figure 9: [Right-time feature] Weather.*

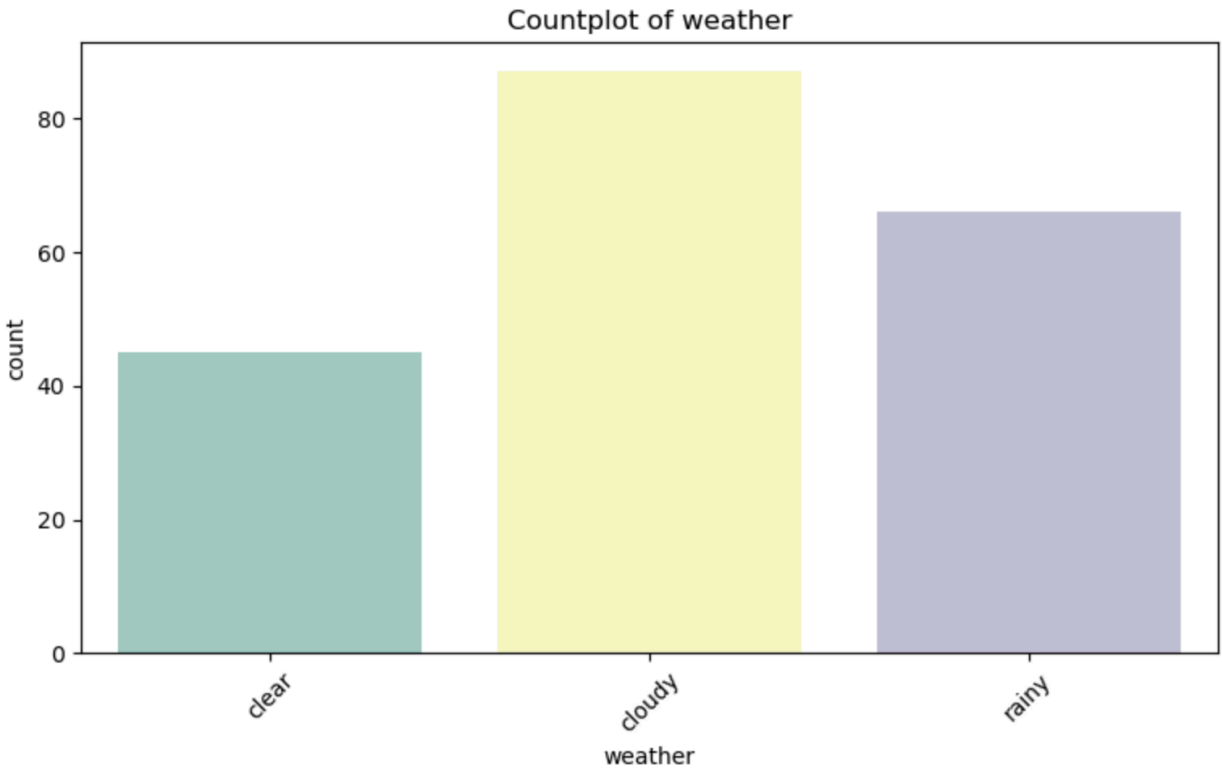

*Figure 10: Target variable distributions for {Meet, Chat, Reject} and {Accept, Reject} configurations.*

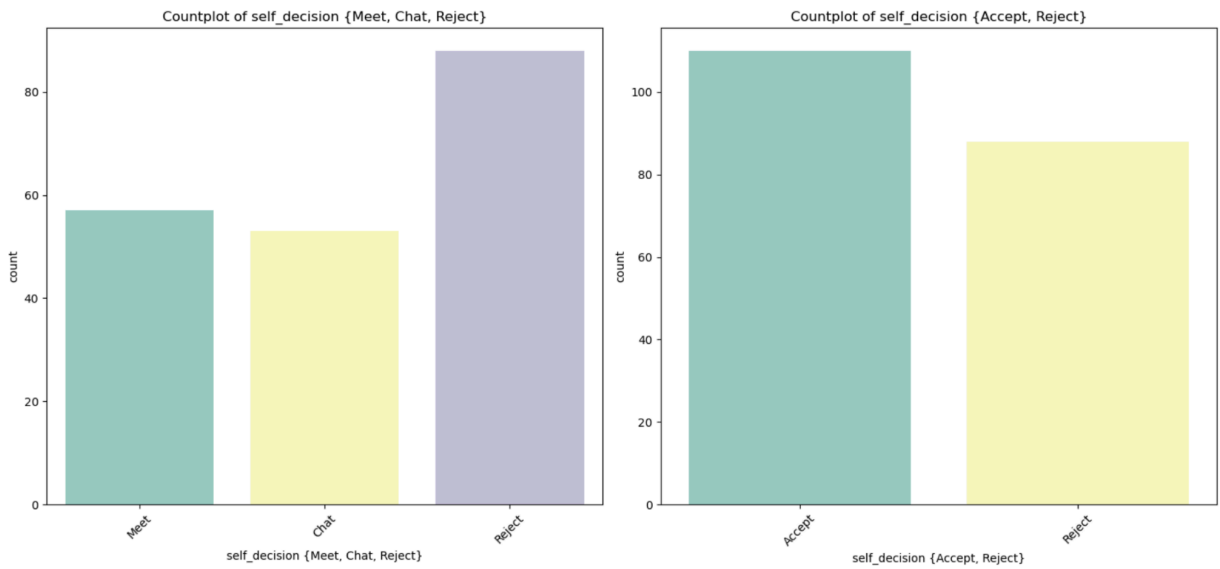

**APPENDIX F:** *Focus group results.*

*Figure 1: Distribution of personality type for individuals participating in the focus group.*

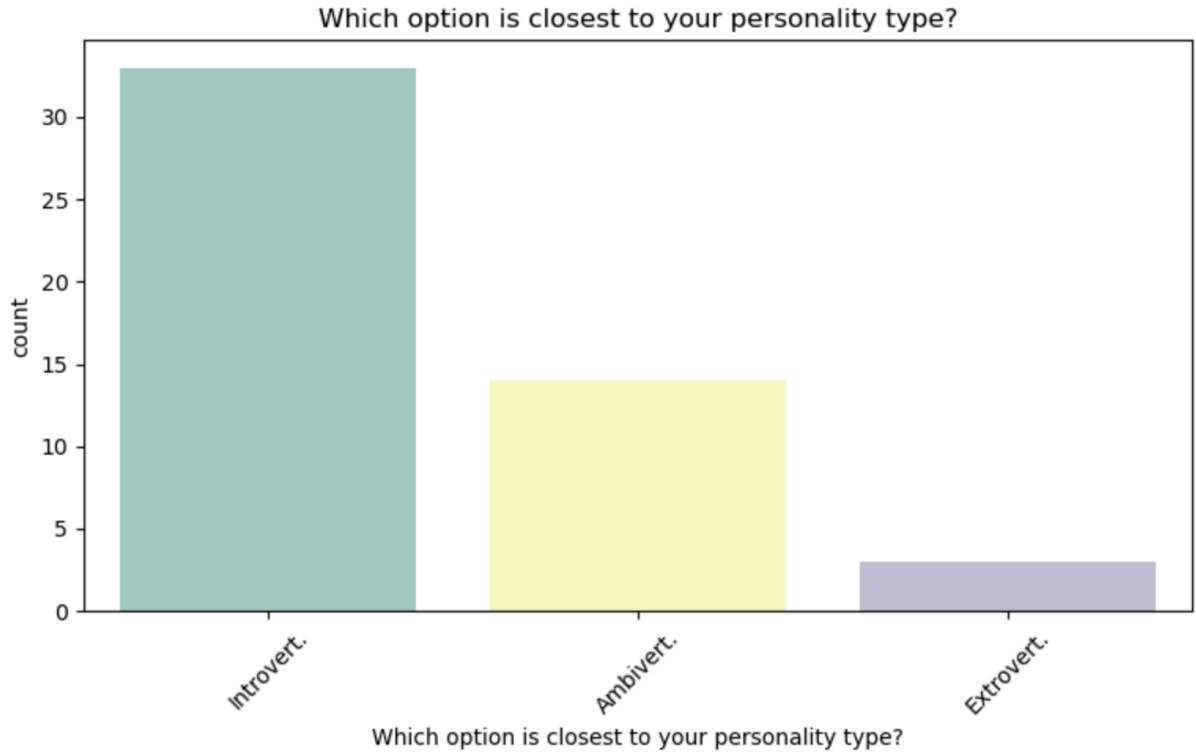

*Figure 2: Questions about frequency and difficulty of social interactions.*

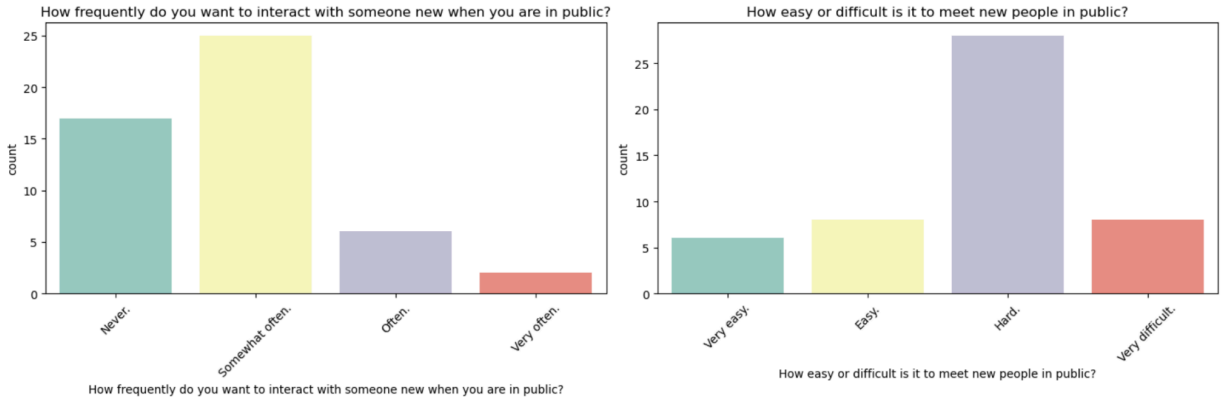

*Figure 3: Questions about social interaction preferences and notifications.*

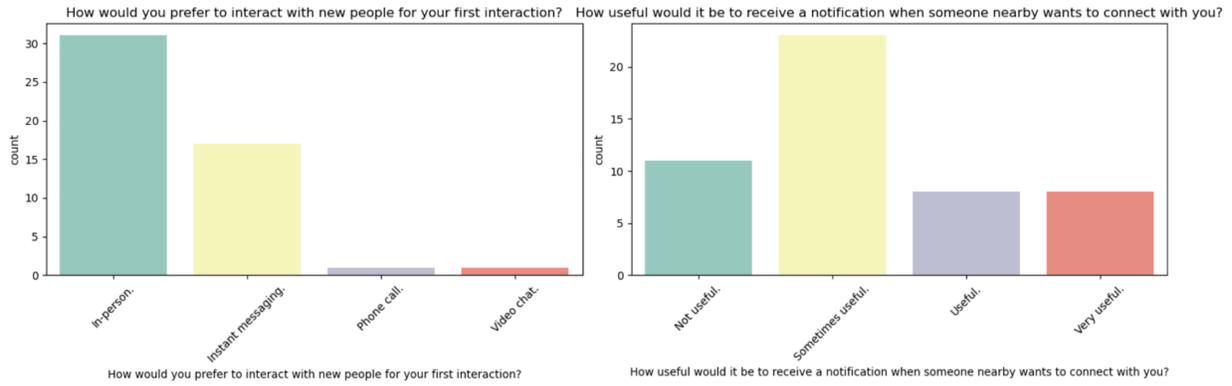

*Figure 4: Questions about music.*

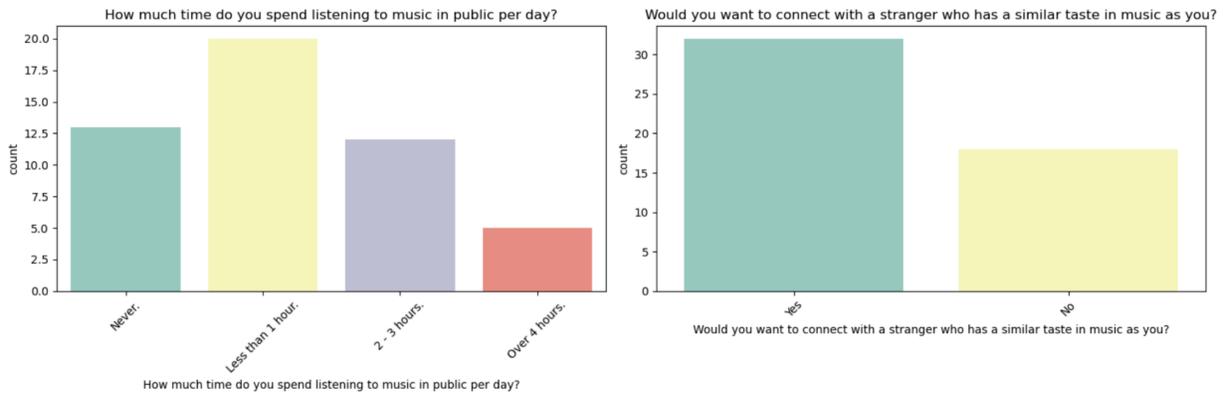

*Figure 5: Personal information sharing preferences.*

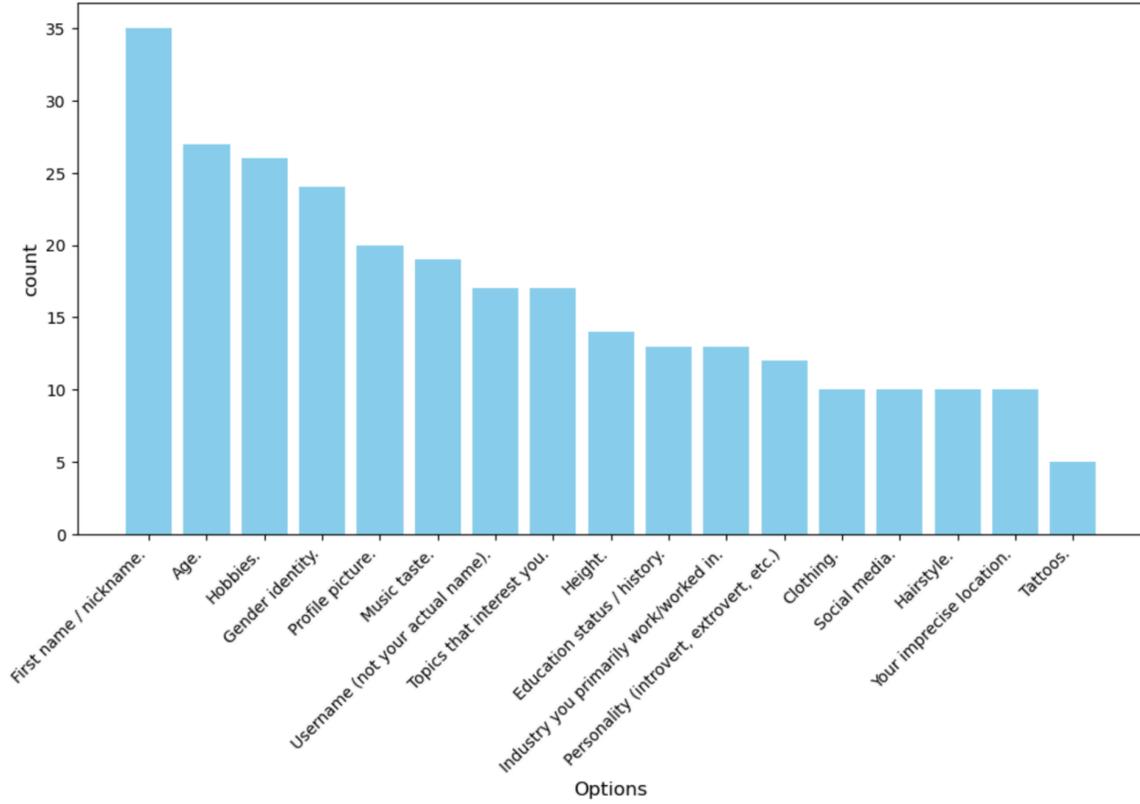

**APPENDIX G:** *GPT-4 prompt and response for understanding user decisions in scenarios.*

*GPT-4 Prompt:*
The following are responses by survey participants responding about another individual. The participants select from the options "Meet in person", "Chat via instant messaging", or "Reject" (denoted in brackets for each response). Then, they also provide an explanation of their choice. Could you summarize the reasons that people choose to "Meet in person", "Chat via instant messaging", or "Reject". The reasons should be ordered by how frequently they are mentioned, and the results should be split up by each response type ("Meet in person", "Chat via instant messaging", and "Reject").

***SURVEY RESPONSES BELOW***
1. [Chat (via instant messaging)] Considering she is an introvert, chatting online would seem to be an easier means of getting to know that person without making it awkward for myself and them.
…
*[additional responses]*
…

*GPT-4 Response:*
Here are the summarized reasons that survey participants chose to "Meet in person", "Chat via instant messaging", or "Reject", ordered by frequency:

Reasons for "Meet in person":
Similar Interests - Shared hobbies, music genres, education, and personality traits.
Age - Similar or close age range.
Proximity - Already in a social environment, workplace, or setting conducive to meeting.
Compatibility - Similar lifestyles, relationship status, and career backgrounds.
Appearance - Tattoos, clothing style, earrings, or similar appearance.

Reasons for "Chat via instant messaging":
Introversion - Easier for introverts to communicate without the pressure of face-to-face interaction.
Group Aspect - Avoiding interruptions in public or with groups of people.
Shared Interests - Hobbies, music genres, or creative arts.
Compatibility - Age, occupation, or education.
Non-Confrontational Approach - Allows participants to gauge compatibility before deciding to meet in person.

Reasons for "Reject":
Age Gap - A significant difference in age or preference for a similar age group.
Social Setting - Being in a sit-down restaurant, workplace, or other setting deemed inappropriate for striking up conversations with strangers.
Group Dynamics - Individuals already engaged with groups of people or appearing too busy.
Lack of Common Ground - Differences in interests, lifestyle, or activities.
Sexual Orientation or Gender - Incompatibility in sexual orientation or preference for specific genders.

**APPENDIX H:** *Parameter grid used to train RandomForestClassifier models.*

```python
from sklearn.ensemble import RandomForestClassifier
from sklearn.model_selection import ParameterGrid

param_grid = {
  'n_estimators': [50, 100, 200, 500],
  'max_depth': [3, 5, 7, None],
  'min_samples_split': [2, 4, 6, 10],
  'min_samples_leaf': [1, 2, 5, 10],
  'max_features': ['sqrt', 'log2', None],
  'min_impurity_decrease': [0, 0.001, 0.01, 0.05, 0.1, 0.2]
}

...

for params in ParameterGrid(param_grid):
  ...

  random_forest = RandomForestClassifier(**params)
  random_forest.fit(X_train, y_train)

  ...
```